\documentclass[conference,final]{IEEEtran}
\IEEEoverridecommandlockouts
\usepackage{cite}
\usepackage{amsmath,amssymb,amsfonts}
\usepackage{algorithmicx}
\usepackage{graphicx}
\usepackage{textcomp}
\usepackage{xcolor}

\usepackage{nicefrac}
\usepackage{siunitx}
\usepackage{array,framed}
\usepackage{booktabs}
\usepackage{
  color,
  float,
  epsfig,
  wrapfig,
  graphics,
  graphicx,
  subcaption
}

\usepackage{textcomp,amssymb}

\usepackage{latexsym,fancyhdr,url}
\usepackage{enumerate}
\usepackage{graphics}
\usepackage{xparse} 
\usepackage{multirow}
\usepackage{csvsimple}
\usepackage{balance}
\usepackage{amsmath, amssymb}
\usepackage{pifont}
\usepackage{graphicx}
\usepackage{textcomp}
\usepackage{xcolor}
\usepackage{booktabs}
\usepackage{soul}
\usepackage{xspace}
\usepackage{mathtools}
\usepackage{wasysym}
\usepackage{algorithm}
\usepackage[noend]{algpseudocode}
\usepackage{kotex}
\usepackage{scalerel,stackengine}
\usepackage{makecell}
\usepackage{fontawesome}
\usepackage{hyperref}

\usepackage{
  tikz,
  pgfplots,
  pgfplotstable
}

\usetikzlibrary{
  shapes.geometric,
  arrows,
  external,
  pgfplots.groupplots,
  matrix
}

\usepackage{mathtools}

\def\BibTeX{{\rm B\kern-.05em{\sc i\kern-.025em b}\kern-.08em
    T\kern-.1667em\lower.7ex\hbox{E}\kern-.125emX}}

\newcommand{\mss}{\xspace{$ms$}\xspace}

\newcommand{\ourtool}{\texttt{MUFFLER}\xspace}

\begin{document}

\title{\ourtool{}: Secure Tor Traffic Obfuscation with Dynamic Connection Shuffling and Splitting}

\author{\IEEEauthorblockN{Minjae Seo\IEEEauthorrefmark{1}, Myoungsung You\IEEEauthorrefmark{2}, Jaehan Kim\IEEEauthorrefmark{2}, Taejune Park\IEEEauthorrefmark{3}, Seungwon Shin\IEEEauthorrefmark{2}, and Jinwoo Kim\IEEEauthorrefmark{4}
}
\thanks{Minjae Seo and Myoungsung You contributed equally to this work.\newline{} Jinwoo Kim is the corresponding author.}
\thanks{This work was partly supported by Institute of Information \& communications Technology Planning \& Evaluation (IITP) grant funded by Korea government (MSIT) (No.~2021-0-00118, RS-2021-II210118: Development of decentralized consensus composition technology for large-scale nodes, 70\%) and the National Research Foundation of Korea (NRF) grant funded by the Korea government (MSIT) (No. 2022R1C1C1006967, 30\%).}
\thanks{This paper is an extended version of our preliminary work~\cite{minjae_seo_2023_8197630}.}

\IEEEauthorblockA{\IEEEauthorrefmark{1}ETRI, Daejeon, Republic of Korea}
\IEEEauthorblockA{\IEEEauthorrefmark{2}School of Electrical Engineering, KAIST, Daejeon, Republic of Korea}
\IEEEauthorblockA{\IEEEauthorrefmark{3}School of Artificial Intelligence, Chonnam National University, Gwangju, Republic of Korea}
\IEEEauthorblockA{\IEEEauthorrefmark{4}School of Software, Kwangwoon University, Seoul, Republic of Korea}}

\maketitle

\begin{abstract}
Tor, a widely utilized privacy network, enables anonymous communication but is vulnerable to flow correlation attacks that deanonymize users by correlating traffic patterns from Tor's ingress and egress segments. Various defenses have been developed to mitigate these attacks; however, they have two critical limitations: (i) significant network overhead during obfuscation and (ii) a lack of dynamic obfuscation for egress segments, exposing traffic patterns to adversaries. In response, we introduce \ourtool{}, a novel connection-level traffic obfuscation system designed to secure Tor egress traffic. It dynamically maps real connections to a distinct set of virtual connections between the final Tor nodes and targeted services, either public or hidden. This approach creates egress traffic patterns fundamentally different from those at ingress segments without adding intentional padding bytes or timing delays. The mapping of real and virtual connections is adjusted in real-time based on ongoing network conditions, thwarting adversaries' efforts to detect egress traffic patterns. Extensive evaluations show that \ourtool{} mitigates powerful correlation attacks with a TPR of 1\% at an FPR of $10^{-2}$ while imposing only a 2.17\% bandwidth overhead. Moreover, it achieves up to 27x lower latency overhead than existing solutions and seamlessly integrates with the current Tor architecture.
\end{abstract}
\section{Introduction}
\label{sec:intro}

The Tor network, one of the most popular privacy networks~\cite{tor_metrics}, provides anonymity for internet users by routing traffic through a global network of volunteer-run relay nodes. Users establish a Tor circuit through three types of relays: entry, middle, and exit nodes. Traffic passes through this circuit to reach the intended services, with each node decrypting one layer of encryption to reveal the subsequent node in the path. This layered encryption ensures that no single node knows both the origin and final destination of the data in transit, providing anonymous and secure communication.

Users of the Tor network can access both \textit{public} and \textit{hidden} services anonymously without revealing their IP addresses. Public services are standard web servers accessible via both Internet and the Tor network. Conversely, hidden services~\cite{dingledine2004tor}, identified by their \emph{.onion} addresses, are only accessible through the Tor network and hide the IP addresses of both users and servers from adversaries. To access hidden services, users first connect to designated rendezvous points to receive access points for these services. Subsequent communications between users and hidden services are routed through circuits connected to the specified access points, ensuring the end-to-end security and anonymity of both users and services.

Given Tor's objective to facilitate anonymous communication for sensitive tasks, it has naturally become a target for various deanonymization attacks. Among these, the flow correlation attack~\cite{murdoch2005low} is particularly powerful, as it correlates unique characteristics of traffic flows, such as packet sizes and interval times, observed from the ingress and egress segments of a Tor connection. In recognition of the threat posed by flow correlation attacks, several obfuscation methods~\cite{obfs4, dyer2012peek, cai2014systematic, juarez2016toward,abusnaina2020dfd,gong2020zero,de2020trafficsliver,nasr2021defeating} have been developed to mitigate them. However, these methods are not well suited to the requirements of the Tor network due to the following two limitations:

\noindent\textbf{Network Inefficiency.} 
The majority of obfuscation methods currently employed to protect the Tor network focus on adding \emph{crafted padding bytes}~\cite{dyer2012peek, cai2014systematic, Yawn2014obfs4, juarez2016toward, gong2020zero, abusnaina2020dfd, nasr2021defeating, scramblesuit} or \emph{inter-packet delays}~\cite{dyer2012peek, cai2014systematic, Yawn2014obfs4, scramblesuit} into packets, thereby hiding the original traffic patterns. 
However, they result in a significant increase in bandwidth consumption, which is a critical concern for the Tor network. Given that Tor relies on limited bandwidth resources provided by volunteers, any addition in bandwidth requirements is \textit{not} trivial~\cite{jansen2019point, yang2015mtor}. Furthermore, intentional packet delays compromise communication latency and user experience, undermining the usability of these mechanisms.

\begin{figure}
    \centering
    \includegraphics[width=\linewidth]{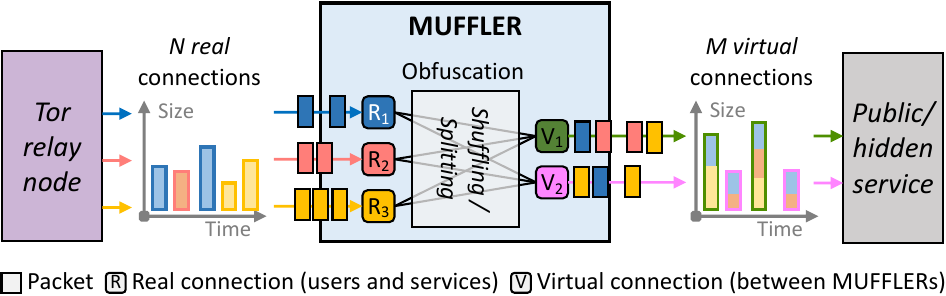}
    \caption{\ourtool{} dynamically obfuscates egress Tor traffic by mapping \textit{N} real connections to \textit{M} virtual connections.}
    \label{fig:odin_intro}
    \vspace{-5mm}
\end{figure}

\noindent\textbf{Lack of Dynamic Obfuscation for Egress Segments.} 
Effective obfuscation for the egress segment is crucial for maintaining anonymity as Tor encryption is fully decrypted when accessing \emph{public} services. However, existing solutions solely focus on obfuscation for the ingress segments, which fails
to conceal critical traffic patterns. Recent machine learning-based attacks~\cite{nasr2018deepcorr,oh2022deepcoffea,guan2023flowtracker} have demonstrated high accuracy in deanonymizing users accessing public services through Tor networks. Similar vulnerabilities are present when accessing Tor \emph{hidden} services. Although final Tor nodes, which are directly connected to hidden services, attempt to enhance anonymity by multiplexing client connections into a single one~\cite{dingledine2004tor}, the multiplexing schema remains static, employing a simple \textit{N:1} traffic pattern\footnote{$N$ represents the number of original connections between users and the hidden service, while $1$ denotes a single multiplexed connection between the guard node and the hidden service.} regardless of network status (e.g., the number of connections). This static nature has enabled a recent machine learning-based attack to correlate users and their destination hidden services with over 99\% accuracy~\cite{lopesflow}. 

\noindent\textbf{Our Approach.}
In this paper, we introduce \ourtool{}, a novel connection-level traffic obfuscation system designed to secure Tor traffic at the egress segment. As shown in Fig.~\ref{fig:odin_intro}, \ourtool{} employs two obfuscation strategies: connection shuffling and connection splitting. These strategies map $N$ real connections (e.g., TCP sessions) from the ingress segment into a set of $M$ virtual connections established between the Tor final node and the target service, whether public or hidden. Packets from each real connection are relayed via a corresponding virtual connection, thereby implementing both $N$:$M$ shuffled and $1$:$M$ split traffic patterns within Tor egress segments. Additionally, \ourtool{} dynamically adjusts the mapping between virtual and real connections in response to changing network conditions (e.g., the number of connections), hindering adversaries’ efforts to identify static mapping patterns. This dynamic connection-level obfuscation creates fundamentally different traffic patterns from those of ingress segments, thereby preventing existing flow correlation attacks and enhancing the anonymity of the Tor network.

Through the implementation of \ourtool{}, we address the aforementioned limitations not tackled by existing solutions~\cite{dyer2012peek, cai2014systematic, Yawn2014obfs4, juarez2016toward, gong2020zero, abusnaina2020dfd, nasr2021defeating, scramblesuit}. It eliminates the need for padding bytes and inter-packet delays, effectively hiding original traffic patterns by dynamically relaying packets received from real connections through virtual connections, achieving an impressively low average bandwidth overhead of only 2.17\%. Additionally, \ourtool{} leverages recent kernel networking features~\cite{eBPF} to avoid unnecessary kernel network stack processes during obfuscation, resulting in 15\% and 24\% improvements in mean and tail latency. Our connection-level obfuscation can seamlessly protect both public and onion services. Extensive evaluations show \ourtool{}'s effectiveness in mitigating powerful flow correlation attacks targeting public and hidden services, achieving a True Positive Rate (TPR) of 1\% at a False Positive Rate (FPR) of $10^{-2}$.

\noindent\textbf{Contributions.} We make the following contributions:
\begin{list}{\labelitemi}{\leftmargin=1em}
\item{The design and implementation of \ourtool{}, a defence system that obfuscates Tor traffic by dynamically mapping real connections into distinct virtual connections without the need for additional padding bytes or per-packet delays.}
\item{The integration of a recent kernel networking feature that reduces kernel network stack processes typically required for obfuscation, thereby decreasing latency overheads.}
\item{Extensive evaluations showing that \ourtool{} prevents several flow correlation attacks, which are not adequately addressed by existing obfuscation methods, while imposing only minimal bandwidth and latency overheads.}
\end{list}

\begin{figure}
    \centering
    \includegraphics[width=.99\linewidth]{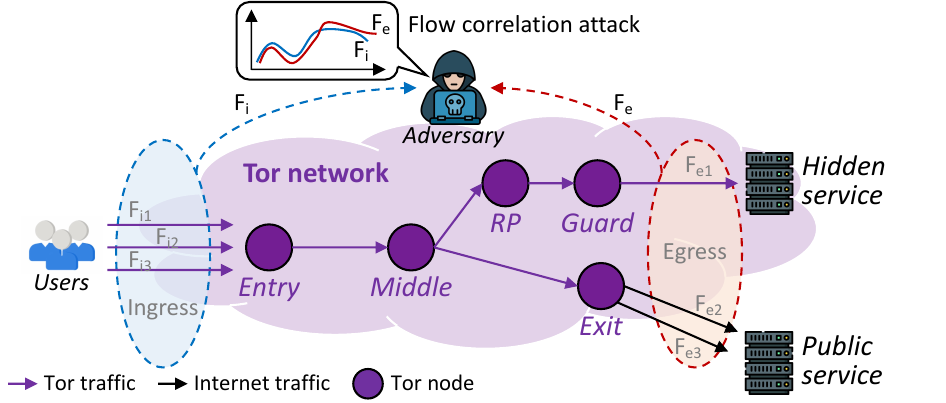}
    \caption{Adversaries can correlate network flow pairs (i.e., ingress flows $F_i$ and egress flows $F_e$) utilizing traffic features.}
    \label{fig:fca}
    \vspace{-4mm}
\end{figure}

\section{Background and Motivation}
\label{sec:background}

\subsection{Flow Correlation Attacks}
In most scenarios of flow correlation attacks, adversaries attempt to link traffic flows (e.g., TCP connections) observed from two points: \textit{ingress flows} ($F_i$ between users and entry nodes) and \textit{egress flows} ($F_e$ between guard/exit nodes and hidden/public services), as shown in Fig.~\ref{fig:fca}. Those two points, with their pair of \emph{associated flows} ($F_i$, $F_e$), carry the information of the real \textit{identity} (i.e., IP address of the user or service), making them attractive targets for adversaries aiming to eavesdrop on Tor traffic. If adversaries collect enough traffic from both points, they can perform flow correlation attacks targeting public services or hidden services, uncovering the IP addresses of users through various traffic analysis techniques.

\noindent\textbf{Flow Correlation Attacks on Public Services.} Initially, several studies have used a statistical metric to measure the flow similarity of \emph{associated flows} ($F_i$, $F_e$) observed from both \textit{ingress} and \textit{egress} points when accessing public services. For example, Sun et al.~\cite{sun2015raptor} used the Spearman correlation coefficient, a nonparametric measure that evaluates the statistical dependence between the rankings of two variables. They extracted \emph{TCP header information}, particularly TCP sequence (SEQ) and acknowledgement (ACK) numbers, from each packet trace to conduct asymmetric correlation analysis, allowing adversaries to observe Tor traffic at both points (\textit{ingress} and \textit{egress}). However, they considered the case where there is a long-lasting Tor connection, requiring a long-term observation, rather than a short-lived Tor connection, which is more common in the real-world Tor network. 

To address the limitations of the statistical metric, recent studies have adopted machine-learning techniques. For example, Nasr et al.~\cite{nasr2018deepcorr} proposed a machine learning-based flow correlation attack, DeepCorr, which can correlate \emph{associated flows} ($F_i$, $F_e$) with high accuracy using short-term observations of Tor connections. Due to the encryption for Tor packets, they leveraged deep neural networks (DNNs) to learn hidden patterns within Tor traffic. They found dominant characteristics in \emph{packet sizes} and \textit{inter-packet delays} when correlating Tor traffic. Their evaluations demonstrated that DeepCorr can correlate associated Tor flows with 96\% accuracy.

\noindent\textbf{Flow Correlation Attacks on Hidden Services.} Recently, a flow correlation attack targeting hidden services has been developed. This attack, known as the SUMo attack~\cite{lopesflow}, leverages a novel technique called the Sliding Subset Sum algorithm to measure flow similarity. The SUMo attack utilizes the absolute packet arrival times to correlate flows, modeling packets received by clients and transmitted by hidden services as bounded time series. This approach allows adversaries to calculate the similarity score of flows by aligning \emph{packet sizes} within sliding windows, enhancing its efficacy in identifying patterns. Although multiple client connections from Tor relay nodes between clients and a hidden service are multiplexed in a single connection, it still exposes the traffic feature of packet sizes in a bounded time series.

\noindent\textbf{Attack Vector.} Flow correlation attacks on Tor, leveraging traffic features such as \emph{TCP header context}, \emph{packet sizes}, and \emph{inter-packet delays}, present a significant privacy concern. This situation highlights an imperative demand for robust defense mechanisms capable of obscuring these attack vectors.

\subsection{Motivation}
\label{sec:motivation}
To prevent adversaries from exploiting the aforementioned attack vectors, various traffic obfuscation solutions~\cite{dyer2012peek, cai2014systematic, Yawn2014obfs4, juarez2016toward, gong2020zero, abusnaina2020dfd, nasr2021defeating} have been developed. These solutions focus on client-side obfuscation, which masks Tor ingress traffic between users (clients) and Tor entry nodes. For example, Tor's official client-side obfuscation methods, such as ScrambleSuit~\cite{scramblesuit} and obfs4~\cite{Yawn2014obfs4}, utilize \emph{padding} and \emph{timing delays} to obscure identifiable features. ScrambleSuit sends MTU-sized packets, adding padding when data is less than the MTU to randomize packet sizes. Obfs4 enhances this approach with stronger cryptographic protections during the initial TLS handshake between users and entry nodes. It employs the ``iat-mode'' (Inter-Arrival Time mode) to adjust traffic timing, offering settings to disable (0) or enable (1 or 2) intentional delays, thus improving obfuscation effectiveness. FRONT~\cite{gong2020zero} is an advanced client-side obfuscation method that hides the ``front'' part of web traffic with random dummy packets.

\begin{figure}[t]
    \centering
    \includegraphics[width=\linewidth]{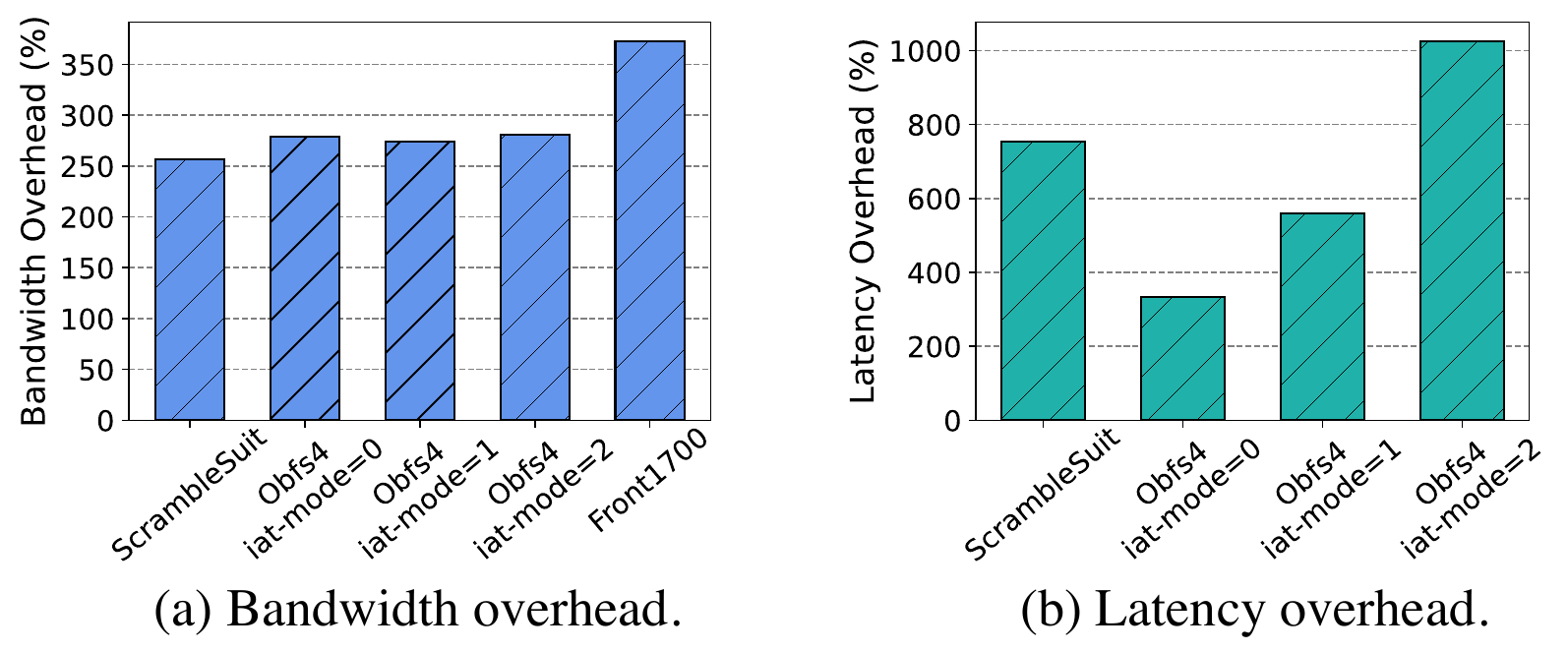}
    \caption{Overheads of existing client-side obfuscation methods.}
    \label{fig:motiv}
    \vspace{-4mm}
\end{figure}

\noindent\textbf{Network Inefficiency.} Despite the enhancements in user protection brought by adopting client-side obfuscation methods, concerns remain about network inefficiencies. To formalize this, we define the bandwidth and latency overhead, partially borrowed from \cite{gong2020zero} as follows: We define a sequence of $n$ packets denoted by $P=\{p_1, p_2, \ldots, p_n\}$, where $p$ is a packet. Each packet $p$ has $(t_i, L_i)$, where $t_i$ is the timestamp and $L_i$ is the length of the $i$-th packet, respectively. We also define $|P|=\sum_{i=1}^{n}L_i$ by the total length of a sequence of packets. Here, let $P$ denote the original sequence and $P'$ the obfuscated sequence after applying a certain obfuscation solution $D$. The bandwidth and latency overhead are defined as follows:
The \textit{bandwidth overhead} $O(D)$ of defense $D$ on $P$ is the total padding length (i.e., increased packet length) divided by the total packet length of the original sequence:
$
    O(D) = \frac{|P'| - |P|}{|P|}
$
The \textit{latency overhead} $T(D)$ of defense $D$ on $P$ is the extra time taken to transmit real packets, divided by the original transmission time. Denoting the timestamp of the last real packet in $P'$ as \( t_k \), then we have:
$
    T(D) = \frac{t_k - t_{n}}{t_{n}}
$

With this definition, we performed a 30-minute web browsing experiment to measure the bandwidth and latency overhead of ScrambleSuit, obfs4 with various iat-mode settings, and FRONT. Fig.~\ref{fig:motiv} (a) shows that these techniques result in approximately a 250\% bandwidth overhead. Additionally, Front1700, which uses a ratio of 1,700 dummy packets for every 10,000 real packets, results in the highest bandwidth overhead, exceeding 350\%. Given the limited bandwidth of the Tor network, any increase in bandwidth consumption is significant. Furthermore, Fig.~\ref{fig:motiv} (b)\footnote{Note that FRONT is excluded from the latency evaluation as it obfuscates collected traffic traces offline rather than real-time traffic.} indicates that ScrambleSuit alone causes a latency overhead of over 700\%. The iat-mode settings further increase latency, with overheads of approximately 300\%, 600\%, and 1,000\% for iat-mode=0, 1, and 2, respectively. These results highlight that existing client-side obfuscation methods introduce substantial bandwidth and latency overhead, exacerbating network performance issues.

\noindent\textbf{Lack of Dynamic Obfuscation for Egress Segments.}
Existing solutions focus on obfuscating traffic at Tor ingress segments (between users and entry nodes). 
However, this client-side obfuscation alone is insufficient for protection against flow correlation attacks, as it fails to hide critical correlation features at Tor egress segments.
This is particularly concerning as all protections provided by Tor networks are stripped at this point, revealing the IP addresses of public services that users connected to.
This limitation enables machine-learning-based attacks~\cite{nasr2018deepcorr, oh2022deepcoffea,guan2023flowtracker} to correlate users with public services with high accuracy. Moreover, hidden services, traditionally considered secure, face similar threats. As shown in Fig.~\ref{fig:fca}. Tor guard nodes multiplex multiple user connections into a single one using a fixed $N$:1 mapping. This schema does not dynamically adjust to changes in network conditions, such as the number of connections, making it ineffective against evolving threats. Such static obfuscation methods allow advanced techniques like SUMo~\cite{lopesflow} to identify correlation features within short time slots, achieving a 99\% accuracy rate in linking users to their destination hidden services.
\section{\ourtool{} Design}
\label{sec:overview}

\subsection{Design Considerations}
To address the aforementioned limitations, we structure \ourtool{} around the following design considerations (DC):

\noindent\textbf{DC1. Efficient Obfuscation without Intentional Overheads.} Existing solutions rely on intentional data padding or timing delays. These methods result in significant bandwidth consumption and increased latency. Thus, the proposed system should obfuscate Tor traffic without intentional overheads while providing the same level of security guarantee.

\noindent\textbf{DC2. Robust and Dynamic Egress Traffic Obfuscation.} 
Existing solutions lack robust server-side obfuscation for the Tor network, particularly for Tor hidden services and the traffic between exit nodes and public services.
This limitation allows adversaries to identify users accessing both hidden and public services. Consequently, the proposed system should implement dynamic server-side obfuscation that comprehensively secures Tor egress traffic directed to both hidden and public services, adapting in real-time to the prevailing network conditions.

\noindent\textbf{DC3. Efficient and Seamless Integration with Tor.}
The compatibility of obfuscation systems with the existing Tor ecosystem is essential for their broad deployment. Therefore, the proposed system should be designed for seamless integration with the Tor ecosystem without requiring modifications to the Tor protocol or the Tor binary. Also, it should obfuscate and process egress traffic in an optimized way to avoid becoming a bottleneck on the Tor network.

\begin{figure}[t]
    \centering
    \includegraphics[width=\linewidth]{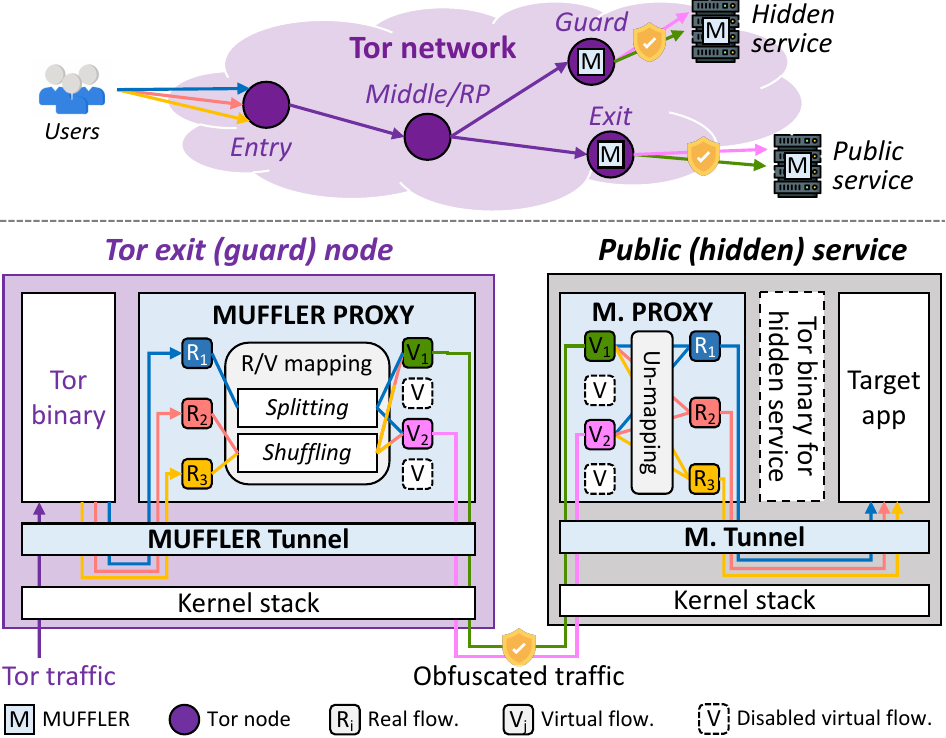}
    \caption{\ourtool{} is deployed between the final Tor relay node and the target service (hidden or public).}
    \label{fig:system_overview}
    \vspace{-3mm}
\end{figure}

\subsection{Threat Model and Assumptions}

\noindent\textbf{Threat Model.} The goal of adversaries is to deanonymize users by correlating a pair of \emph{associated flows} ($F_i$, $F_e$), including \textit{ingress} flows ($F_i$) and \textit{egress} flows ($F_e$), as shown in Fig.~\ref{fig:fca}. They compare several unique characteristics of Tor traffic (e.g., packet sizes and timings) instead of attempting to decrypt the content of Tor traffic. To achieve their goal, we consider adversaries who have capabilities to eavesdrop on Tor traffic (i.e., ingress and egress flows) passively. 

We believe this scenario is practical in the real world. In recent years, many threat actors are running hundreds of malicious Tor nodes in order to
intercept Tor traffic~\cite{wright2002analysis, jansen2018inside, borisov2007denial}. Adversaries can further leverage BGP hijacking attacks, which becomes increasingly frequent, 
by a means to redirect the Tor traffic to themselves~\cite{sun2015raptor}. There might also be more powerful adversaries, such as governmental agencies, who can perform wiretapping attacks on multiple Internet ASes/IXPs~\cite{feamster2004location, murdoch2007sampled, johnson2013users, nithyanand2016measuring, sun2015raptor} or intercontinental fiber optics~\cite{wiretapping}.

\noindent\textbf{Deployment Scenario.} 
We assume a scenario where users seek anonymous access to public or hidden (onion) services via Tor networks to circumvent censorship and Internet tracking. To protect these users from flow correlation attacks, we propose that \ourtool{} be deployed on machines operating target services—public or hidden—and on the \textit{egress Tor nodes} directly connected to these services. For hidden services, the egress nodes are defined as Tor guard nodes, whereas for public services, they are Tor exit nodes. This deployment model is practical, as operators of hidden services and users can designate a set of predefined guard nodes through their Tor configuration files (\texttt{torrc})~\cite{tor_rc_entry}.
Furthermore, considering the pivotal role of exit nodes in forwarding Tor traffic to the Internet, several reputable organizations and privacy-conscious entities already manage trusted exit nodes~\cite{DuckDuckGo_exit, kim2018sgx} equipped with enhanced security features. \ourtool{} can be effectively installed on these exit nodes to offer additional protection for users accessing security-enhanced public services.

\subsection{\ourtool{} Overview}
Fig.~\ref{fig:system_overview} shows the overall architecture of \ourtool{} comprising two main components: the \ourtool{} \texttt{Tunnel} and the \ourtool{} \texttt{PROXY}. These components are adeptly deployed on target servers (either hidden or public servers) and an egress Tor relay node directly connected to these servers. This deployment strategy ensures the transparent and efficient obfuscation of egress traffic. The operation of \ourtool{} is structured into two main phases:

\noindent\textbf{Obfuscation Phase.} This phase begins with the Tor binary on the Tor egress node processing incoming Tor traffic from its previous relay node.
To achieve transparent obfuscation, the \ourtool{} \texttt{PROXY} runs as a reverse proxy, intercepting traffic from the Tor binary and sending obfuscated traffic to target services.
When doing so, this proxy handles non-onion traffic for securing connections between exit nodes and public services, as well as onion traffic for protecting connections between guard nodes and hidden services. 
While the Tor binary and the \ourtool{} \texttt{PROXY} run on the same host, the traditional local-host communication can lead to multiple packet copies and repetitive kernel network stack operations.
Thus, the \ourtool{} \texttt{Tunnel} facilitates socket-level packet redirection. Data transmitted from the Tor binary's socket is directly inserted into the corresponding socket of the \ourtool{} \texttt{PROXY}. This redirection operates before packetization, avoiding extra packet copies and kernel stack operations. 

As shown in Fig.~\ref{fig:system_overview}, the \ourtool{} employs connection-level obfuscation by mapping real TCP connections—those between users and services—to a set of virtual connections established in conjunction with its paired proxy on the server side. This approach allows packets from real connections to be transmitted through corresponding virtual connections. Here, \ourtool{} utilizes two novel mapping strategies:
(i)~connection shuffling and (ii)~connection splitting. In connection shuffling mode, packets from \textit{N} real connections are shuffled across \textit{M} virtual connections, thereby creating merged traffic patterns.
In contrast, connection splitting mode spreads packets from a single real connection over \textit{M} multiple virtual connections, creating evenly distributed traffic patterns.
Moreover, the mapping between real and virtual connections dynamically adjusts during runtime based on each connection's state, further hindering adversaries from discovering correlation patterns.

\noindent\textbf{De-obfuscation Phase.} On the server side, the obfuscated packets within these virtual connections are restored to their original real connections. 
The \ourtool{} \texttt{PROXIES} on both the egress node and the target server synchronize their connection mapping states through our control protocol. Based on these mapping states, shuffled packets are de-obfuscated to their corresponding real connections. Similarly, packets distributed across various virtual connections are consolidated into a single real connection. These de-obfuscation processes maintain data integrity for server processing. The \ourtool{} \texttt{PROXY} subsequently forwards de-obfuscated packets to the intended service application (e.g., web services). For hidden services, de-obfuscated packets are relayed back to the Tor binary for final onion processing before reaching the application. Throughout these transmission steps, the \ourtool{} \texttt{Tunnel} ensures direct data transfer from 
source to destination sockets, effectively bypassing unnecessary kernel stack involvements.

By implementing the above processes, \ourtool{} addresses the three design considerations. 
Specifically, it maps real connections to a set of virtual connections, which exhibit distinct patterns, such as variations in the number of connections and SEQ/ACK numbers. This connection-level obfuscation effectively conceals the patterns of real connections without relying on padding bytes or timing delays (\textbf{DC1}). Furthermore, \ourtool{} runs between the Tor egress nodes and their target services, protecting Tor egress traffic destined for both public and hidden services, thus mitigating the risk of correlation attacks in the egress segments (\textbf{DC2}). Additionally, our proxy-based design enables seamless integration with the existing Tor ecosystem without any modifications. The \ourtool{} \texttt{Tunnel} complements this design by avoiding the overheads associated with traditional proxy-based obfuscations (\textbf{DC3}).

\subsection{\ourtool{} Tunnel}
\ourtool{} adopts a proxy-based approach that intercepts outgoing packets from the Tor binary. Although this design aligns transparently with the Tor binary, it introduces several inefficiencies during packet redirection to the \ourtool{} \texttt{PROXY}. Specifically, this redirection incurs two packet copies and two kernel stack processes, which consume host system resources and contribute to increased latency~\cite{qi2022spright, qi2023middlenet}. To address these issues, \ourtool{} employs the \ourtool{} \texttt{Tunnel} that enables traffic redirection at the socket level between the Tor binary and the \ourtool{} \texttt{PROXY}, thereby avoiding these overheads, as shown in Fig.~\ref{fig:system_overview}.

The \ourtool{} \texttt{Tunnel} runs on the kernel space and monitors socket system calls (e.g., \textit{connect} and \textit{send}) invoked by the Tor binary. When the Tor binary attempts to create a socket with the target service, this tunnel intercepts the relevant system call. It then records the original destination of the socket and modifies the system call's parameter, making the Tor binary to establish a socket with the \ourtool{} \texttt{PROXY} instead of the target service. When the Tor binary later attempts to send a new data through the created socket, this tunnel intercepts the data and redirects it from the source socket’s Tx queue to the destination socket’s Rx queue, thus avoiding the traditional kernel stack operations and unnecessary packet copies. After obfuscation, the \ourtool{} \texttt{PROXY} utilizes the recorded destination information to send the obfuscated traffic to the original destination through corresponding virtual connections, ensuring proper routing to the target services.


\begin{figure}[t]
    \centering
    \includegraphics[width=\linewidth]{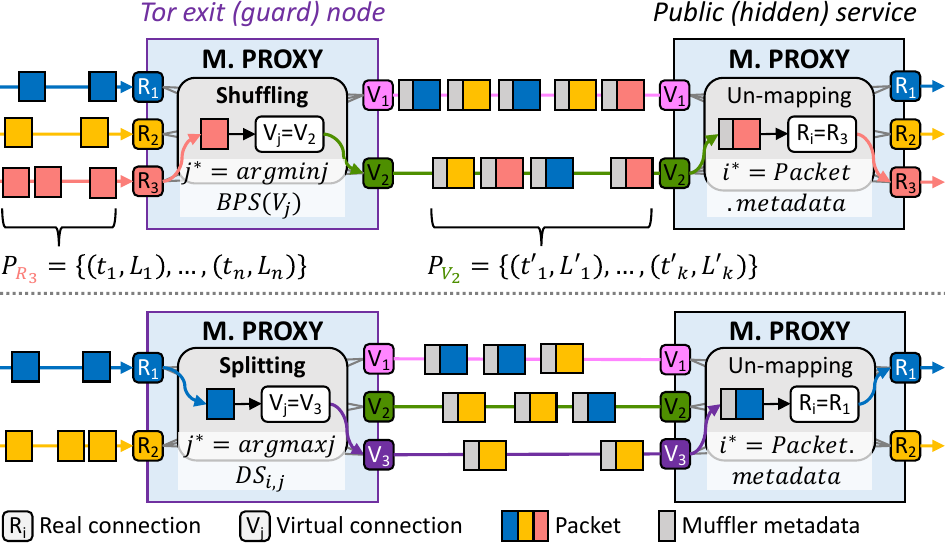}
    \caption{An overview of connection shuffling (upper) and connection splitting (lower) of the \ourtool{} \texttt{PROXIE}s.}
    \label{fig:shuffle}
    \vspace{-4mm}
\end{figure}

\subsection{Obfuscation with Connection Shuffling and Splitting}
\label{sec:virtual}

To ensure robust security for Tor egress traffic, \ourtool{} maps real connections to virtual connections through connection-level obfuscation.
As shown in Fig.~\ref{fig:shuffle}, it selects a proper mapping strategy among the following two strategies based on the current number of real connections (\textit{N}).

\noindent \textbf{Connection Shuffling.} When \textit{N} exceeds a shuffle threshold ($S$)---a sufficient volume for effective shuffling---the \ourtool{} \texttt{PROXY} runs in connection shuffling mode. This mode adjusts the number of active virtual connections (\textit{M}) to meet the following schema:
$M = \min\left(\left\lfloor \alpha N \right\rfloor, M_{\text{min}}\right)$ where $\alpha$ is a shuffling factor less than one, ensuring a lower number of virtual connections than real connections, and $M_{\text{min}}$ is the minimum number of virtual connections maintained for robust obfuscation. 
As shown in the upper part of Fig.~\ref{fig:shuffle}, packets from \textit{N} real connections are relayed through \textit{M} virtual connections, creating an $N:M$ shuffled traffic pattern where each virtual connection receives a mix of packets from multiple real connections. Also, the selection of a $j$-th virtual connection for each packet considers the current bytes per second (BPS) of each virtual connection using the following schema: $j^* = \arg \min_j BPS(V_j)$ where $BPS(V_j)$ indicates the BPS of a virtual connection. This schema ensures that incoming packets are distributed evenly across virtual connections, preventing any single virtual connection from becoming a bandwidth bottleneck and revealing a unique pattern. 

\noindent \textbf{Connection Splitting.} Although connection shuffling effectively hides the correlation features of real connections, it becomes less effective with a smaller number of real connections and is infeasible with only one real connection. To address this limitation, the \ourtool{} \texttt{PROXY} switches to connection splitting mode when \textit{N} falls below the shuffling threshold ($S$). In this mode, it activates new virtual connections according to the following schema: $M = \max(\beta N, M_{\text{min}})$
where $\beta$ is a splitting factor greater than one, designed to increase the spread of traffic across a greater number of virtual connections.
As shown in the lower part of Fig.~\ref{fig:shuffle}, this mode splits packets from a real connection evenly across the \textit{M} virtual connections, creating $1:M$ split traffic patterns. In addition, the \ourtool{} \texttt{PROXY} leverages a dissimilarity score to optimize traffic distribution: $DS_{i,j} = |BPS(R_i) - BPS(V_j)|$ where $BPS(R_i)$ and $BPS(V_j)$ represent the BPS of $i$-th real and $j$-th virtual connections, respectively. 
The selected virtual connection for packet relay maximizes this dissimilarity score: $j^* = \arg \max_j DS_{i,j}$, ensuring that the traffic distribution across virtual connections does not reflect the bandwidth patterns of the real connection. This connection splitting mode hides individual correlation patterns from adversaries even when real connections are limited.

Our connection-level obfuscation generates correlation features that are fundamentally distinct from those observed at the Tor ingress segments, including differences in TCP context, packet sizes, and inter-packet delays. As shown in the upper part of Fig.~\ref{fig:shuffle}, individual packets within the packet sequences of three real connections ($P_{R_1}$, $P_{R_2}$, $P_{R_3}$) are multiplexed into two virtual connections ($P_{V_1}$, $P_{V_2}$), resulting in obfuscated packet sequences. Assume that adversaries observe a specific correlation feature such as BPS for the obfuscated sequence $P_{V_2}$. Consequently, $BPS(P_{V_2})$ is calculated as $|P_{V_2}|/(t'_k-t'_1)$, where $k$ denotes the last packet index of the obfuscated sequence $P_{V_2}$. It becomes clear that $BPS(P_{V_2})$ is not equivalent to $BPS(P_{R_1})$, $BPS(P_{R_2})$, or $BPS(P_{R_3})$, thereby rendering adversaries' flow correlation attacks ineffective. 


\noindent
\textbf{Connection Un-mapping.}
To maintain the integrity of service processing, the \ourtool{} \texttt{PROXY} at the target service should correctly dispatch packets from virtual connections back to their corresponding original real connections (i.e., un-mapping). The \ourtool{} \texttt{PROXIES} on both ends facilitate this process by exchanging control commands through their virtual connection. In the current design, a control command is 8-byte long, including four 2-byte fields: a command type field, two operand fields, and one reserved field. When relaying a data packet from a real connection, a \textit{relay} type command is utilized. As shown in Fig.~\ref{fig:shuffle}, this command is embedded within each packet's payload as metadata, including real connection IDs. The receiving proxy utilizes this information to accurately dispatch packet data from each virtual connection to the appropriate sockets linked with their original real connections. Additionally, control commands are used for synchronizing configurations between \ourtool{} \texttt{PROXIES}. For instance, the creation of a new virtual connection triggers the exchange of a \textit{create} type command between proxies prior to executing TCP handshakes. A \textit{keep-alive} type command is also used to maintain deactivated virtual connections.
Note that since \textit{relay} commands are embedded into the actual data packets, they may introduce some bandwidth overhead. However, with each \textit{relay} command adding only 8 bytes, this overhead is considered negligible, when compared to previous data padding methods~\cite{gong2020zero, Yawn2014obfs4} that add between 500 to 1,000 dummy bytes per packet to create fixed-length traffic patterns.
\section{Evaluation}
\label{sec:eval}
In this section, we first evaluate the bandwidth and latency overhead associated with \ourtool{} compared to existing solutions. Next, we evaluate the effectiveness of \ourtool{} in obfuscating Tor traffic against several flow correlation attacks.

\subsection{Prototype Implementation}
We have developed a full prototype of the \ourtool{} \texttt{PROXY}, leveraging the core functionality of HAProxy~\cite{haproxy} and extending it using the Go language. The \ourtool{} \texttt{PROXY} consists of client and server components. These components initiate multiple long-lived TLS (or TCP) connections, referred to as base connections, which are utilized to create virtual connections. To facilitate the division of a single base connection into multiple virtual connections, we implemented a set of control commands: \textit{create}, \textit{remove}, \textit{relay}, and \textit{keep-alive}, as described in Section~\ref{sec:virtual}. 
Additionally, the \ourtool{} \texttt{Tunnel} leverages three types of eBPF programs~\cite{tran2019making, yang2024network}. These programs are attached to the Tor binary and the \ourtool{} \texttt{PROXY} to monitor socket system calls, modify system call arguments, store socket descriptors, and redirect data from source sockets to destination sockets.

\subsection{Experimental Environment}
\noindent\textbf{Private Tor Testbed.} To evaluate the feasibility of \ourtool{} while mitigating potential impacts on real-world Tor users, we create a private Tor testbed atop a Kubernetes cluster, utilizing three physical machines. The testbed includes a number of Tor relay nodes—entry, middle, exit nodes, and directory authorities. Each machine is equipped with two Intel Xeon Silver 4114 CPUs and 64GB of memory. This configuration ensures an isolated environment distinct from the real Tor network, facilitating a controlled comparison of performance between \ourtool{} and existing studies. Moreover, to further isolate resource usage and minimize interference, we run public and hidden services on separate servers equipped with an Intel i9-10900X CPU with 256GB of memory.

\noindent\textbf{Defense Settings.} 
To evaluate the effectiveness of \ourtool{} within our testbed, we configure its parameters as follows. We set the shuffling threshold ($S$) to four, meaning that \ourtool{} runs on the connection shuffling mode when $N$ exceeds four. Additionally, we configure the settings of $\alpha$, $\beta$, and $M_{\text{min}}$ (see Section~\ref{sec:virtual}) to 0.1, 2, and 3, respectively, to optimize performance and obfuscation effectiveness. As \ourtool{} is the first egress-side obfuscation for Tor, we compare it with the following two popular ingress-side obfuscation systems. We first consider obfs4 across different inter-arrival time modes (iat-mode  0, 1, and 2), which manage the inter-packet delays. FRONT~\cite{gong2020zero} is a state-of-the-art system that obfuscates the initial (front) part of web traffic traces using dummy padding data. We consider two variants of FRONT—FRONT 1700 and FRONT 2500—where the numbers represent the number of dummy packets per 10,000 real packets, as specified in the original paper~\cite{gong2020zero}. As the specific size of the dummy packets used in FRONT is not directly stated in the paper, we set the size of dummy data to MTU size (1,500 bytes), which is a prevalent setting in existing studies~\cite{meier2022ditto, dyer2012peek}.

\begin{figure}[t]
  \centering
  \includegraphics[width=\linewidth]{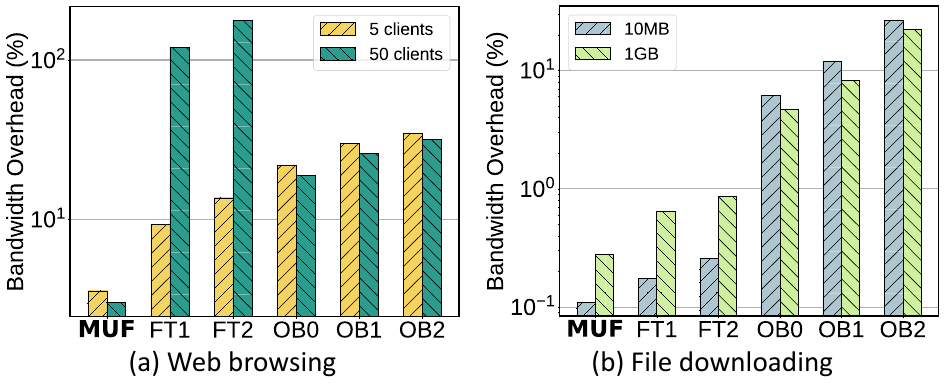}
  \caption{Bandwidth overhead of obfuscation solutions. \textit{FT1} and \textit{FT2} indicate FRONT 1700 and FRONT 2500. \textit{OB0-2} represents obfs4 with iat-mode of 0, 1, and 2, respectively.}
  \label{fig:bandwidth_eval}
  \vspace{-5mm}
\end{figure}

\subsection{Bandwidth Overhead}
We evaluate bandwidth overhead using the same definitions presented in Section~\ref{sec:motivation}. We run two types of applications commonly used on Tor networks: website browsing and file downloading, as shown in Fig.~\ref{fig:bandwidth_eval}.

\noindent\textbf{Website Browsing.}
In this experiment, we analyze website browsing performance using two widely used microservice applications~\cite{SockShop, RobotShop}, simulating an environment with 5 clients and 50 clients to mimic realistic scenarios. Fig.~\ref{fig:bandwidth_eval}~(a) shows the evaluation results. \ourtool{} employs the connection shuffling mode across virtual connections, which introduces only minimal overheads of 3.56\% and 3.01\% under 5 and 50 clients, respectively. This overhead consists of our \textit{relay} commands embedded into the payload of each packet. In contrast, existing solutions show significant overheads due to their padding-based obfuscation strategies. FRONT in its 1700 and 2500 settings shows significant bandwidth overheads of 9\% and 13\% under 5 clients. This overhead further increases to exceeding 100\% when handling 50 clients. In addition, obfs4  under different iat-modes shows a stepwise escalation in bandwidth usage for both 5 clients and 50 clients. Specifically, with 5 clients, the bandwidth usage increases to 21.84\%, 29.93\%, and 34.61\%, respectively, for each iat mode.

\noindent\textbf{File Downloading.}
In scenarios involving file downloads, characterized by long-lasting HTTP connections and large-sized packets, \ourtool{} demonstrates exceptionally low bandwidth overheads, below 1\%, as shown in Fig.~\ref{fig:bandwidth_eval}~(b). This minimal overhead stems from the fact that the size of our relay commands is significantly smaller than the overall data packet size during such downloads. FRONT exhibits slightly higher overheads than \ourtool{}, under 2\%, because it only adds padding data to the initial part of the HTTP connection. On the other hand, obfs4 experiences significantly higher bandwidth overheads across all file sizes due to its method of inserting substantial amounts of dummy data to obscure traffic patterns. Despite handling large-sized packets, obfs4’s approach of extensive padding results in consistent and considerable overheads—averaging 5.41\% for iat-mode~0, 10.13\% for iat-mode~1, and 24.27\% for iat-mode~2.

\subsection{Latency Overhead}
\begin{figure}[t]
    \centering
    \includegraphics[width=0.99\columnwidth]{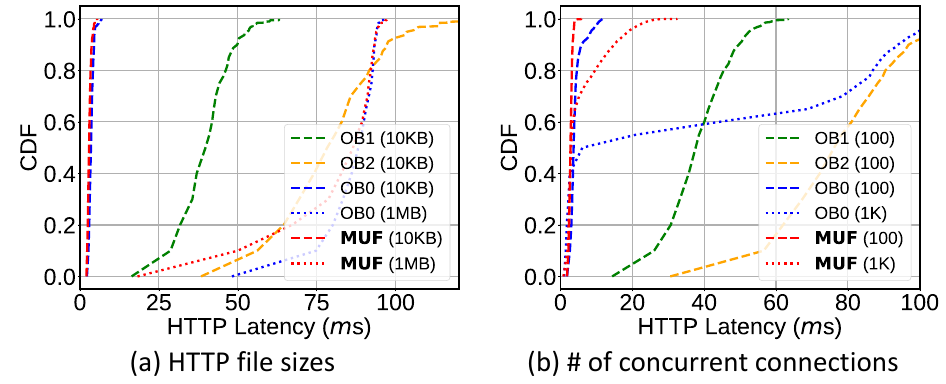}
    \caption{The web browsing latency measurements of \ourtool{} with the comparison of existing obfuscation methods.}
    \label{fig:latency_eval}
    \vspace{-3mm}
\end{figure}

\begin{figure*}[t]
  \centering
  \includegraphics[width=\linewidth]{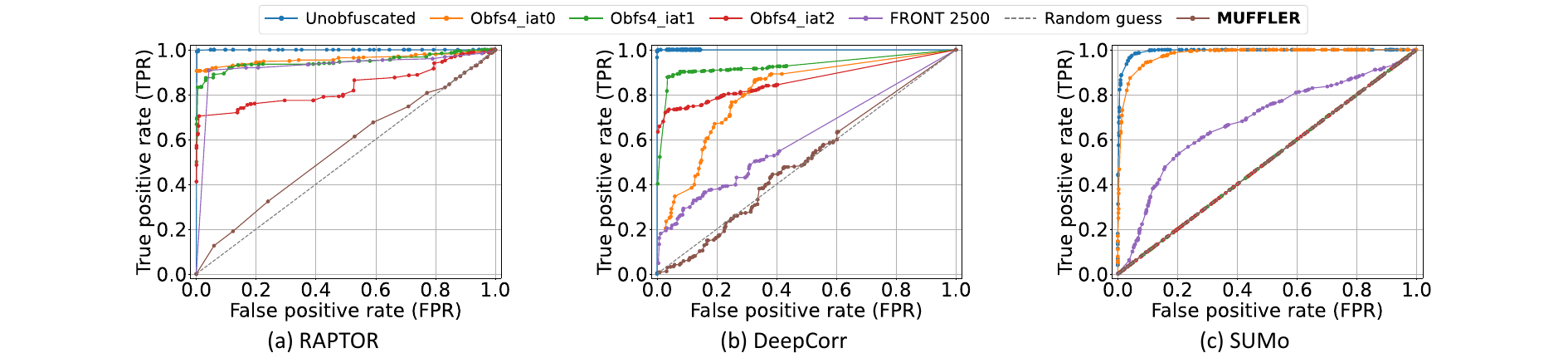}
  \caption{The ROC curves of the performance of the flow correlation attacks across obfuscation methods.}
  \label{fig:obfuscation_effectiveness}
  \vspace{-4mm}
\end{figure*}

Here, we evaluate the latency overhead introduced by \ourtool{}. Using the same testing environment as in previous evaluations, we measure HTTP round trip latency from the moment the Tor binary within the exit node sends an HTTP request to when it receives the HTTP response from the public web service. For a fair comparison, obfs4 is configured to operate between the Tor exit node and the target service, mirroring \ourtool{}'s setup. Note that FRONT is excluded from this evaluation as it obfuscates already collected traffic traces offline, rather than real-time traffic during run-time.

Fig.~\ref{fig:latency_eval}~(a) shows the evaluation results for varying HTTP response sizes. When handling 10KB HTTP traffic, \ourtool{} achieves a mean latency of 2.887 \mss, outperforming obfs4-iat2 by 27x. In addition, compared to obfs4-iat0, which incurs minimal delay at the cost of degraded security, \ourtool{} demonstrates a 17.1\% reduction in latency. These performance improvements become more evident with larger HTTP traffic sizes. When handling 1MB of traffic, \ourtool{} significantly outperforms all other solutions by a significant margin. Fig.~\ref{fig:latency_eval} (b) shows the evaluation results for different levels of concurrent connections with 10KB HTTP files. While previous solutions struggle under high connection loads due to their reliance on inefficient obfuscation methods, \ourtool{} outperforms other solutions by up to 27x, achieving a mean latency of 3.4\mss and a P90 latency of 14.9\mss under 1K connections.

\subsection{Obfuscation Effectiveness}
\label{sec:effectiveness}
We evaluate the security effectiveness of \ourtool{} against several powerful flow correlation attacks within our private Tor network, comparing it to previous obfuscation methods across two scenarios: accessing public services and hidden services. In the first scenario, we employ RAPTOR~\cite{sun2015raptor}—a statistical metric-based attack, and DeepCorr~\cite{nasr2018deepcorr}—a machine learning-based attack. We collect 500 ingress and egress flows during web browsing to train the DeepCorr model and another set of 500 flows for testing RAPTOR and DeepCorr. During traffic collection, we use various numbers of concurrent flows from 2 to 100. Note that the training for DeepCorr is performed exclusively on non-obfuscated flows to establish a baseline, while the evaluation phase for both attacks uses obfuscated flows\footnote{We conduct an additional experiment training and evaluating the DeepCorr model on obfuscated flows, but it shows only a negligible F1-score increase of 0.004 due to MUFFLER’s dynamic shuffling and splitting. Thus, we omit this evaluation, focusing instead on a more realistic adversarial setting.}.
The hyperparameters for these models are selected according to the original studies to ensure optimal attack performance. The second scenario focuses on hidden services, where we employ SUMo~\cite{lopesflow}—an advanced flow correlation attack specifically designed to identify users accessing Tor hidden services. We deploy multiple hidden services to our private Tor network by leveraging popular microservice web applications~\cite{RobotShop, SockShop}, simulating realistic hidden web service traffic. We collect 500 ingress/egress flows when accessing these hidden services with various concurrent connections ranging from 2 to 100 to generate test traffic. 

To quantify the effectiveness of the flow correlation attacks, we measure the True Positive Rate (TPR) and False Positive Rate (FPR) of the adversaries' predictions. A \emph{true positive} is identified when the adversaries correctly correlate an actual correlated flow pair, while a \emph{false positive} is identified when two irrelevant flows are mistakenly predicted as correlated. In evaluating \ourtool{}, each flow at the egress segment (virtual connection) contains multiplexed packets, making it unsuitable to directly match the egress flow with its correlated ingress flow (real connection). For a fair comparison of \ourtool{} with previous solutions, we define the ground truth for a correlated flow pair (ingress and egress) based on the \textit{averaged flow similarity}. Specifically, the similarity of a flow pair is calculated by the Euclidean distance in packet size and inter-packet delay, averaged within a specific time window. We consider an ingress flow (real connection) and an egress flow (virtual connection) are correlated if their pair exhibits the highest averaged flow similarity.

\noindent\textbf{Public Services.} 
Fig.~\ref{fig:obfuscation_effectiveness}~(a) shows the effectiveness of various obfuscation methods against sophisticated flow correlation attacks. In the absence of obfuscation, RAPTOR and DeepCorr exhibit high TPRs of 99\% and 100\%, respectively, at an FPR of $10^{-2}$. \ourtool{}, however, demonstrates exceptional defense capabilities; against RAPTOR, depicted in Fig.~\ref{fig:obfuscation_effectiveness}~(a), it achieves a TPR of 0\%, effectively mitigating this statistical metric-based attack. This result underscores the robustness of \ourtool{} in concealing critical TCP sequence and acknowledgment numbers by rerouting packets through distinct virtual connection contexts. Against the DeepCorr attack, which leverages machine learning techniques, \ourtool{} outperforms baseline methods, including FRONT2500, obfs4-iat0, obfs4-iat1, and obfs4-iat2, each of which achieves TPRs of 9\%, 26\%, 77\%, and 64\%, respectively. \ourtool{} maintains a TPR of only 1\% at the same FPR level. This superior performance is attributed to \ourtool{}'s dynamic adjustment of mapping between real and virtual connections, effectively shuffling and splitting traffic features like BPS, PPS, and inter-packet delays across various virtual connections. Such dynamic connection-level obfuscation hinders DeepCorr's ability to match flow-level features, thereby defending against both statistical and machine learning-based flow correlation attacks.

\noindent\textbf{Hidden Services.} 
In the context of hidden services, SUMo, designed to correlate traffic flows to hidden services, achieves a TPR of 88\% without obfuscation, as shown in Fig.~\ref{fig:obfuscation_effectiveness}~(c). 
While obfs4-iat0 allows a significant correlation with a TPR of 64.8\%, it performs markedly better under other configurations like obfs4-iat1 and obfs4-iat2, where SUMo is less effective. This ineffectiveness arises because SUMo is specifically designed to de-multiplex merged traffic flows through sliding subset sums—a technique that struggles against the randomly padded and delayed traffic patterns introduced by methods like obfs4 and FRONT~\cite{lopesflow}. Note that \ourtool{} mainly runs in connection splitting mode during this evaluation as the Tor guard node merges multiple flows into single one. While the number of real connections is limited, \ourtool{} severely diminishes SUMo’s effectiveness by splitting packets from each real connection across multiple virtual connections, considering the individual BPS patterns of each connection. This leads to a drastic reduction in SUMo’s TPR to 1\% at an FPR of $10^{-2}$, rendering its accuracy to levels comparable to random guessing. \ourtool{}’s effectiveness stems from its dynamic connection splitting mechanism that spreads packets to different virtual connection sets and adjusts the mapping in run-time, preventing SUMo from identifying consistent features for sliding subset sum calculations.

\begin{figure}[t]
    \centering
    \includegraphics[width=0.99\columnwidth]{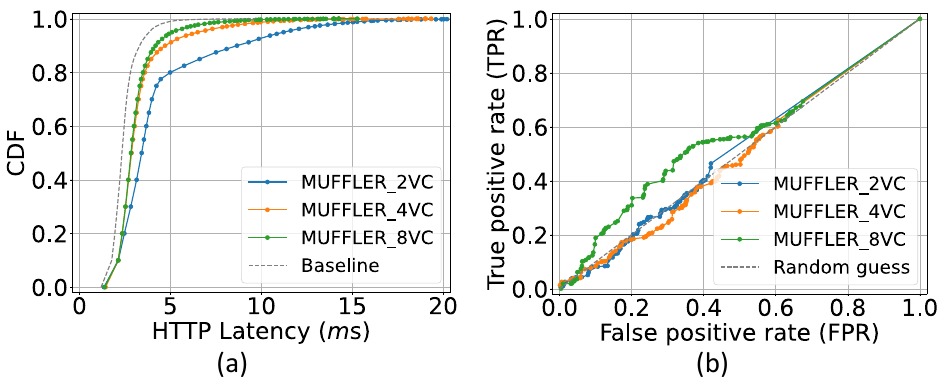}
    \caption{The latency and obfuscation effectiveness under different shuffling factors. The baseline means direct 1:1 mapping.}
    \label{fig:shuffling_factor}
    \vspace{-4mm}
\end{figure}

\subsection{Impact of Shuffling Factor}
We further assess the impact of varying the shuffling factor ($\alpha$) on \ourtool{}'s performance against flow correlation attacks and the associated overhead. For this, we employ 100 concurrent real connections and map them to 2, 4, and 8 virtual connections using different shuffling factors (i.e., $\alpha=0.02$, $0.04$, and $0.08$). We focus on measuring HTTP latency and the security effectiveness against the advanced attack, DeepCorr.

As shown in Fig.~\ref{fig:shuffling_factor} (a), the latency overhead increases linearly with the degree of shuffling. Specifically, in 2 virtual connections setting, which offers the most robust security, \ourtool{} shows a mean latency of 3.69\mss and P90 latency of 8.66\mss. These values represent 49.1\% and 161\% increases in overhead, respectively, compared to the baseline that uses one-to-one mapping without security. This overhead is caused by virtual connections becoming bottlenecks under high degrees of shuffling, resulting in packet drops. In the 8 virtual connections setting, as shown in Fig.~\ref{fig:shuffling_factor} (b), there is a slight reduction in security effectiveness, achieving a TPR of 1.04\% at an FPR of $10^{-2}$ against DeepCorr, but it still performs comparably to random guessing. Remarkably, the latency overhead in this scenario is marginal, showing approximately 0.5\mss for mean latency and 0.8\mss for P90 latency. This evaluation clearly demonstrates that 
\ourtool{} can dynamically adjust the number of virtual connections according to network conditions to balance security needs with performance, making it a viable solution for enhanced Tor traffic obfuscation. Note that we exclude a similar evaluation for the splitting factor ($\beta$), as it shows analogous results.

\section{Related Work}
\label{sec:related_work}


Packet-level obfuscation utilizes strategies such as padding and timing delays to standardize packet rates and timings, thereby concealing exploitable patterns for correlation attacks. Studies like BuFLO~\cite{dyer2012peek} and CS-BuFLO~\cite{cai2014cs} transmit fixed-length packets at regular intervals to standardize the appearance of traffic, while obfs4~\cite{Yawn2014obfs4} employs encryption and padding to render Tor traffic indistinguishable from regular internet traffic. Similarly, WTF-PAD~\cite{juarez2016toward} adapts padding based on observed outgoing traffic patterns in Tor, inserting dummy messages to obscure statistically unlikely packet delays. FRONT~\cite{gong2020zero} obfuscates the initial parts of traffic traces by adding dummy packets at the middle relay in the Tor network, which are subsequently removed before reaching the destination, ensuring that the packet count and distribution are randomized for each trace. Surakav~\cite{gong2022surakav} adopts a machine learning-based approach to obfuscate original traffic patterns by mimicking the patterns of different applications using a GAN-based pattern generator. While these methods enhance privacy, BuFLO and CS-BuFLO can significantly increase bandwidth usage, which is challenging for a network reliant on volunteer resources. Moreover, the need for Tor protocol modifications to remove dummy packets, along with FRONT’s non-inline operation, presents deployment challenges. Surakav offers diverse defense strategies but still incurs high network overhead due to padding and timing delays.

Alternative approaches adopt routing-level obfuscation. For example, TrafficSliver~\cite{de2020trafficsliver} distributes TCP traffic across multiple Tor circuits with unique entry points to prevent flow analysis attacks, but also necessitate Tor protocol modifications, posing deployment hurdles within the existing Tor architecture. Recent methods such as DFD~\cite{abusnaina2020dfd} and BLANKET~\cite{nasr2021defeating} employ adversarial examples to counteract deep learning-based fingerprinting attacks~\cite{seo2022heimdallr, seo2024gshock} by disrupting recognizable traffic patterns or creating adversarial perturbations. However, these techniques require detailed knowledge of the adversaries' DNN models and hyperparameters, which limits their practical application and widespread use in real-world scenarios.
\section{Conclusion and Future Work}
\label{sec:conclusion}
We introduce \ourtool{}, a novel obfuscation system to protect Tor egress traffic from flow correlation attacks. By dynamically shuffling and splitting packets from real connections to a set of virtual connections, \ourtool{} conceals correlation features effectively without the need for padding data or timing delays. Evaluations show that \ourtool{} prevents state-of-the-art flow correlation attacks, achieving a TPR of 1\% at an FPR of $10^{-2}$, while imposing only a 2.4\% bandwidth overhead and 27x lower latency overhead than existing solutions. 

While \ourtool{} effectively prevents flow correlation attacks on Tor, its robustness can be improved.
For example, when there is a single real connection between the final Tor node and the destination service, the distribution of packets across virtual connections may reveal patterns to adversaries. Future work could address this by fragmenting packets in the real connection into smaller segments, embedding reconstruction metadata in segmented packets, and splitting them as new packets across multiple virtual connections. This approach prevents adversaries from identifying original traffic patterns from virtual connections, even when there is a single real connection. Although this method involves extra headers and metadata for segmented packets, it remains more efficient than existing solutions that rely on hundreds of bytes of padding.

\bibliographystyle{IEEEtran}
\bibliography{main}

\end{document}